\documentclass[superscriptaddress,pra,aps,twocolumn,10pt,longbibliography]{revtex4-1}
\usepackage[utf8]{inputenc}
\usepackage[T1]{fontenc}
\usepackage[english]{babel}
\usepackage{bm}
\usepackage{amsmath}
\usepackage{bbm}
\usepackage{amsfonts}
\usepackage{graphicx}
\usepackage[caption=false]{subfig}
\usepackage{natbib}
\usepackage{hyperref}
\usepackage{cleveref}
\usepackage{multirow}
\usepackage{booktabs}
\usepackage{epstopdf}
\usepackage[dvipsnames]{xcolor}
\usepackage{braket}
\usepackage{float}
\usepackage{soul}
\usepackage[boxed]{algorithm}
\usepackage{algpseudocode}
\hypersetup{
    colorlinks=true,       
    linkcolor=cyan,          
    citecolor=magenta,        
    filecolor=magenta,      
    urlcolor=cyan,           
    runcolor=cyan
}

\newcommand{\beq}{\begin{eqnarray}}
\newcommand{\eeq}{\end{eqnarray}}
\newcommand{\bes} {\begin{subequations}}
\newcommand{\ees} {\end{subequations}}
\newcommand{\Tr}{\mathrm{Tr}}
\newcommand{\tr}{\mathrm{Tr}}

\newcommand{\ignore}[1]{}

\graphicspath{{figs/}}

\begin{document}

\title{Augmented fidelities for single qubit gates}

\author{Filip Wudarski}
\email{filip.a.wudarski@nasa.gov}
\author{Jeffrey Marshall}
\email{jeffrey.s.marshall@nasa.gov}

\affiliation{QuAIL, NASA Ames Research Center, Moffett Field, California 94035, USA}
\affiliation{USRA Research Institute for Advanced Computer Science, Mountain View, California 94043, USA}

\author{Andre Petukhov}
\affiliation{QuAIL, NASA Ames Research Center, Moffett Field, California 94035, USA}
\affiliation{Google Inc., Santa Barbara, California 93117, USA}
\altaffiliation[Permanent]{ address}

\author{Eleanor Rieffel}
\affiliation{QuAIL, NASA Ames Research Center, Moffett Field, California 94035, USA}

\begin{abstract}
An average gate fidelity is a standard performance metric to quantify deviation between an ideal unitary gate transformation and its realistic experimental implementation. The average is taken with respect to states uniformly distributed over the full Hilbert space. We analytically (single-qubit) and numerically (two-qubit) show how this average changes if the uniform distribution condition is relaxed, replaced by parametrized distributions -- {\it polar cap} and von Mises-Fisher distributions -- and how the resulting fidelities can differentiate certain noise models. 
In particular, we demonstrate that Pauli channels with different noise rates along the three axes can be faithfully distinguished using these augmented fidelities.
\end{abstract}

\maketitle

\section{Introduction}
Impressive progress in quantum technologies has taken quantum computing from a theoretical framework to an experimental playground, where basic proof-of-principle concepts can be tested and verified. The most prominent recent example of the latter is the demonstration of quantum advantage \cite{aruteQuantumSupremacyUsing2019b}: that even the imperfect currently available quantum hardware can perform tasks intractable for the most powerful supercomputers. Further advances toward more capable and robust  
quantum hardware depend on gaining a better understanding of underlying physical effects, including the characterization of noise in actual quantum hardware.

In theory, quantum process tomography (QPT) \cite{MohseniQPT2008} can be used to exhaustively benchmark a quantum device, identifying all of its imperfections. QPT reconstructs the full process matrix $\chi$ (of size $2^n\times 2^n$, where $n$ is the number of qubits), a matrix that encodes complete information about underlying quantum transformation (including unwanted effects caused by noise). Unfortunately, QPT scales 
exponentially with system size, becoming impractical for systems larger than
a few qubits \cite{Weinstein_2004}. Intuitively it seems plausible that well-controlled systems will have only few dominating error sources - i.e. the $\chi$ matrix will be sparse up to some accuracy. Therefore, 
lower parameter approximations and associated metrics and protocols that could assess the performance of quantum devices, and identify the crucial elements of $\chi$, are promising approaches to noise characterization and quantum hardware benchmarking. Currently, the most common
figures of merit considered
are: diamond norm \cite{Benenti_2010}, minimum fidelity \cite{lu2020direct} or average fidelity \cite{bowdreyFidelitySingleQubit2002,nielsenSimpleFormulaAverage2002,horodeckiGeneralTeleportationChannel1999}. All three techniques yield a single value that characterizes deviations from the ideal transformation. A parameter count shows that all three methods 
provide only limited information about the process matrix for a device. 
Nevertheless, they are all valuable benchmarking tools that allow researchers to capture and quantify some of the most relevant 
aspects of the behavior of quantum devices and their building blocks - qubits and gates.  

Special attention should be given to the average fidelity - currently the figure of merit of performance metric. It describes an error between an ideal and experimental realization of a gate, and is averaged over all possible states uniformly distributed in the Hilbert space (according to the Haar measure). Unlike the diamond norm and the min fidelity,  the average gate fidelity 
can be efficiently estimated through protocols like randomized benchmarking (RB) \cite{emersonScalableNoiseEstimation2005,erhardCharacterizingLargescaleQuantum2019a,magesanRobustRandomizedBenchmarking2011,wallmanRandomizedBenchmarkingConfidence2014}, cross-entropy benchmark \cite{boixoCharacterizingQuantumSupremacy2018,aruteQuantumSupremacyUsing2019b} or direct fidelity estimation \cite{flammiaDirectFidelityEstimation2011,dasilvaPracticalCharacterizationQuantum2011,moussaPracticalExperimentalCertification2012}.
However, being only a single parameter the metric  
cannot distinguish various noise models; it reports only a single element of the process matrix, the $\chi_{00}$ element, which is associated with the depolarizing rate. That is, it effectively identifies all channels as depolarizing channels.

In this article, we propose to relax the uniform distribution condition and introduce augmented fidelity metrics via parametrized distributions. In particular, we analytically investigate what information about noise processes can be extracted from average fidelity with respect to a von Mises-Fisher distribution (a normal distribution in directional statistics) and a {\it polar cap} distribution, i.e. a uniform distribution over a subset of states parametrized by polar angle (colatitude). This approach augments standard uniform-average fidelity by adding extra tunable parameters to the metric, that depend on distribution properties. 
This work provides a partial solution to the problem posed by Nielsen in \cite{nielsenSimpleFormulaAverage2002} 
regarding gate fidelities over non-uniform distributions. Additionally, we derive the maximal spread in fidelity the value (the difference between minimum and maximum attainable values)  and provide error bars (based on the standard deviation derived from the considered distributions) for processes that share the same depolarizing rate. In particular, we show how to identify noise biases \cite{TuckettPRLUltrahigh2018,TuckettPRXTailoring2019}  in Pauli channels. Our analysis 
focuses mainly on single qubit gates, where analytical formulas are derived, but we also open the discussion for similar approaches for multi-qubit gates, in particular we numerically show local distribution for two-qubit gates and how they differ from the uniform-average fidelity.

\section{Average Gate Fidelity}
Consider the fidelity between a state transformed according to a given unitary (gate) transformation (ideal action) $U$ and the same state transformed with noisy realization of $U$, which is a completely positive and trace preserving (CPTP) map  $\mathcal{E}_U$. 
The average fidelity of the noisy realization is the average of this fidelity over all initial (pure) states distributed uniformly in the entire Hilbert space. 
Bowdrey et al. \cite{bowdreyFidelitySingleQubit2002} introduced a simple formula for calculating average gate fidelity for single qubit gate, which was later generalized to multi-qubit gates and connected with entanglement fidelity \cite{nielsenSimpleFormulaAverage2002,horodeckiGeneralTeleportationChannel1999}.  The average gate fidelity (from now on referred as ``uniform-average fidelity'') for $n$-qubit gates is therefore given by
{{\small
\begin{equation}\label{eq:avg_fid}
\bar{F}(U, \mathcal{E}_U) = \bar{F}(\mathcal{U}^\dag\circ \mathcal{E}_U) =  \frac{\sum_{k=0}^{2^{2n}-1} \Tr\Big(U V_k^\dag U^\dag\mathcal{E}_U(V_k)\Big)+2^{2n}}{2^{2n}(2^n+1)},
\end{equation}
}}where $V_k$ are traceless unitary matrices forming an orthonormal basis with respect to Hilbert-Schmidt inner product ($\Tr(V_kV_j^\dag) = 2^n \delta_{kj}$,  $\Tr(V_k) = 0$ for $k=1,\ldots, 2^{2n}-1$ and $V_0= \mathbbm{1}$). 
By writing the composed map
\begin{equation}\label{eq:map}
\mathcal{E}(\rho) \equiv \mathcal{U}^\dag\circ\mathcal{E}_U(\rho) =\sum_{k,l=0}^{2^{2n}-1}\chi_{kl}V_k\rho V_l^\dag, 
\end{equation}
where the $\chi$ matrix is called the process matrix for $\mathcal{E}$,
it can be demonstrated that the average fidelity only depends on the $\chi_{00}$ element corresponding to a ``depolarizing'' rate, a unitary invariant element, i.e.
\begin{equation}\label{eq:chi00}
 \bar{F}(\mathcal{U}^\dag\circ\mathcal{E}_U) =  \frac{2^n\chi_{00}+1}{2^n+1}.
 \end{equation}
The above formula demonstrates inability to distinguish different noise processes that differ in other $\chi$ parameters, and this 
limitation stems from the averaging procedure and properties of the Haar measure. In order to have a more sensitive metric, we propose to use several different initial state distributions. In addition to averaging over all pure states distributed uniformly, we 
explore two models: i) uniform distribution parametrized by a polar angle $\Theta\in[0,\pi]$, which we call {\it polar cap} distribution (e.g. for $\Theta=\pi/2$ we have a distribution over the northern hemisphere, while for $\Theta=\pi$ we recover the entire space distribution), and ii) von Mises-Fisher distribution around a state $\ket{\psi}$ (without loss of generality we can fix it to $\ket{0}$) parametrized by ``variance'' parameter $\kappa$.  From now on we will focus only on single qubit gates, and will refer to the investigated fidelities as {\it{augmented}} fidelities,
leaving extensions to multi-qubit systems to later work.

\subsection{Polar Cap Distribution}
First let us define a single state gate fidelity as
\begin{equation}\label{eq:single_state_fidelity}
    F_{|\psi\rangle\langle\psi|}(U,\mathcal{E}_U) = \tr\Big(U|\psi\rangle\langle\psi| U^\dag\mathcal{E}_U(|\psi\rangle\langle\psi|) \Big),
\end{equation}
where $U$ is the investigated gate, $\mathcal{E}_U$ its CPTP (imperfect) realization and $\ket{\psi}$ is a state upon which the gate acts. Since measuring and computing Eq.~(\ref{eq:single_state_fidelity}) for all possible states is infeasible, one usually reports the average fidelity value, which is taken over $\ket{\psi}$ distributed uniformly in the entire Hilbert space.

Following derivation from \cite{bowdreyFidelitySingleQubit2002} we define the restricted average gate fidelity $\bar{F}_\Theta(U,\mathcal{E}_U)\equiv \bar{F}_\Theta$ as
\begin{eqnarray}\label{eq:rest_fid}
\bar{F}_\Theta &= &\int F_{\ket{\psi}\bra{\psi}}(U,\mathcal{E}_U) d\Omega = \nonumber\\
&=&\frac{1}{S(\Theta)} \int_{\theta=0}^\Theta\int_{\phi=0}^{2\pi}\Tr\Big(U\big[\sum_{j=0}^3 c_j(\theta,\phi)\frac{\sigma_j}{2} \big]U^\dag\times\nonumber \\
&\times&\mathcal{E}_U\big[ \sum_{k=0}^3 c_k(\theta,\phi)\frac{\sigma_k}{2}\big] \Big)\sin\theta d\phi d\theta,
\end{eqnarray}
where $S(\Theta):=2\pi (1-\cos \Theta)$ is the solid angle for normalization of the distribution, and $c_j(\theta,\phi)$ are pure state's Bloch vector coefficients: $c_0(\theta,\phi) = 1,  c_1(\theta,\phi) = \sin\theta\cos\phi, c_2(\theta,\phi)=\sin\theta\sin\phi$ and $c_3(\theta,\phi)=\cos\theta$.  
These correspond respectively to the identity matrix $\sigma_0$, and the three $x,y,z$ Pauli matrices $\sigma_{1,2,3}$. 
Note that Eq.~(\ref{eq:single_state_fidelity}) assumes coordinate system where $\theta=0$ corresponds to $\ket{0}$ state, and the polar cap distribution is centered around it. However, transformation to an arbitrary central state is straightforward via rotation $\ket{\tilde{\psi}}=U_R\ket{\psi}$. Now performing the integration leaves us with 
\begin{eqnarray}
    \bar{F}_\Theta&= & \frac{1}{2} + \frac{(2+\cos \Theta) \sin^2 \frac{\Theta}{2}}{12}\sum_{k=1}^2  \tr \Big(U \sigma_kU^\dag \mathcal{E}_U \big[\sigma_k\big]\Big) + \nonumber\\
    &+& \frac{1+\cos \Theta + \cos^2 \Theta}{12}\tr \Big(U \sigma_3U^\dag \mathcal{E}_U\big[ \sigma_3\big]\Big)+ \nonumber\\
   & +& \frac{1+\cos \Theta}{8}  \tr \Big(U \sigma_3U^\dag \mathcal{E}_U\big[ \sigma_0\big]\Big).\label{eq:fid_theta}
\end{eqnarray}
It is transparent that for $\Theta = \pi$ one recovers result Eq.~(\ref{eq:avg_fid}) for uniform distribution over the entire Hilbert space.
Expressing the composed gate-noisy-gate map $\mathcal{E}=\mathcal{U}^\dag\circ \mathcal{E}_U$ in the form Eq.~(\ref{eq:map}),  one can show that
\begin{eqnarray}
    \frac{1}{4}\sum_{k=1}^2\tr \Big(\sigma_k \mathcal{E} \big[\sigma_k\big]\Big) & = &\chi_{0,0}-\chi_{3,3},\label{eq:chi_el1}\\
   \frac{1}{2} \tr \Big(\sigma_3 \mathcal{E} \big[\sigma_3\big] \Big)& = &\chi_{0,0}-\chi_{1,1}-\chi_{2,2}+\chi_{3,3},\label{eq:chi_el2}\\
    \frac{1}{8}\tr \Big(\sigma_3 \mathcal{E} \big[\sigma_0\big]\Big) & = & \mathrm{Re}(\chi_{0,3}),\label{eq:chi_el3}
\end{eqnarray}
where we used the properties of $\chi$ matrix that guarantee the CPTP condition, in particular $\mathrm{Re}(\chi_{0,3}) = - \mathrm{Im}(\chi_{1,2})$. The above equations (\ref{eq:fid_theta}-\ref{eq:chi_el3}) are correct for distributions centered around the North Pole (i.e. state $\ket{0}$). However, if the center is selected to be one of $\sigma_1$ or $\sigma_2$ eigenstates, then Eqs.~(\ref{eq:fid_theta}-\ref{eq:chi_el3}) will experience a permutation $3\leftrightarrow1$ or $3\leftrightarrow2$, respectively. For more generic central state the polar cap average  fidelity $\bar{F}_\Theta$ will in principle depend non-trivially on all $\chi$ matrix entries, apart from the imaginary parts of $\chi_{0,k}$ for $k=1,2,3$.

\subsection{von Mises-Fisher Distribution}
\begin{figure}
\center
\includegraphics[width = 0.2\textwidth]{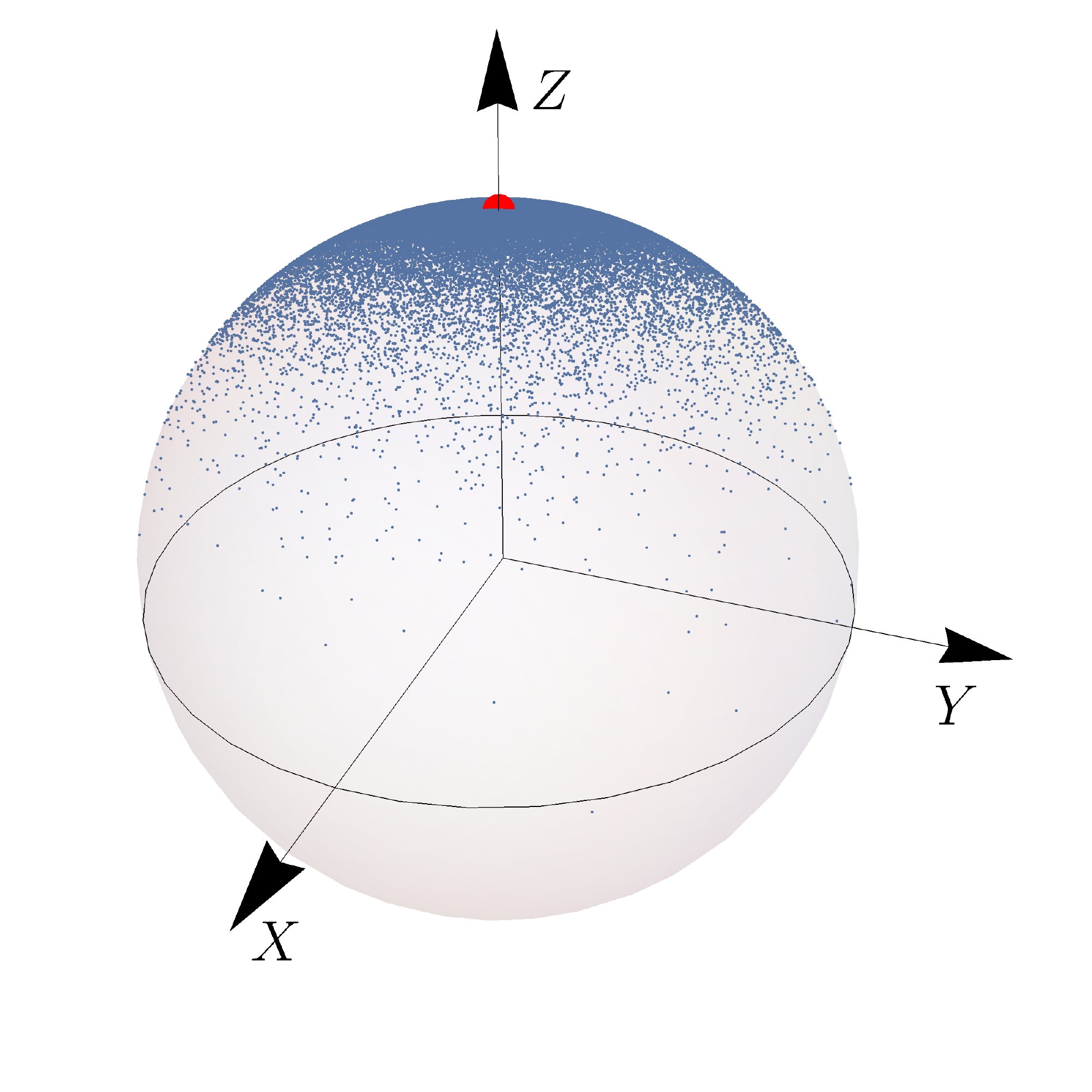}
\includegraphics[width = 0.2\textwidth]{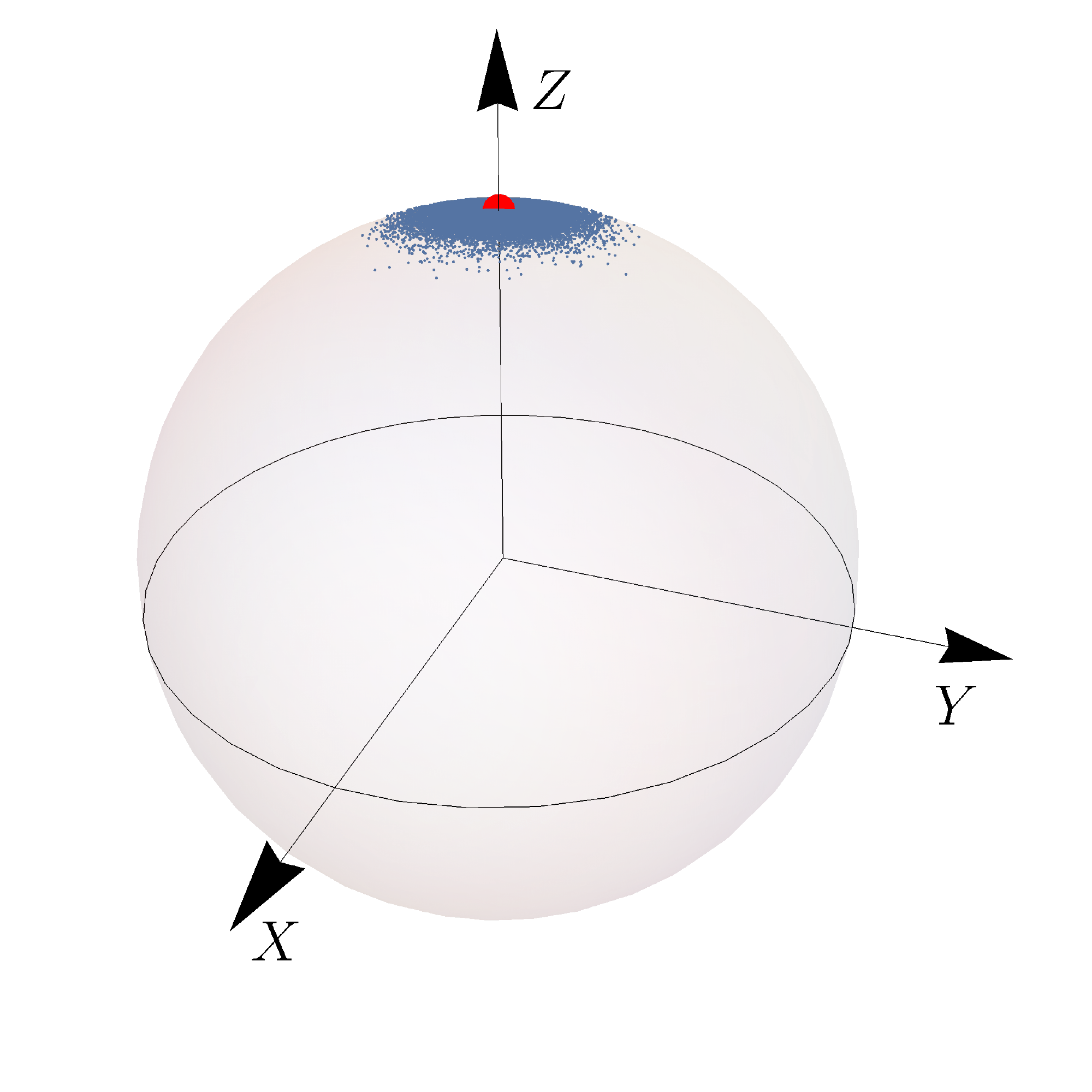}\\
\includegraphics[width=0.4\textwidth]{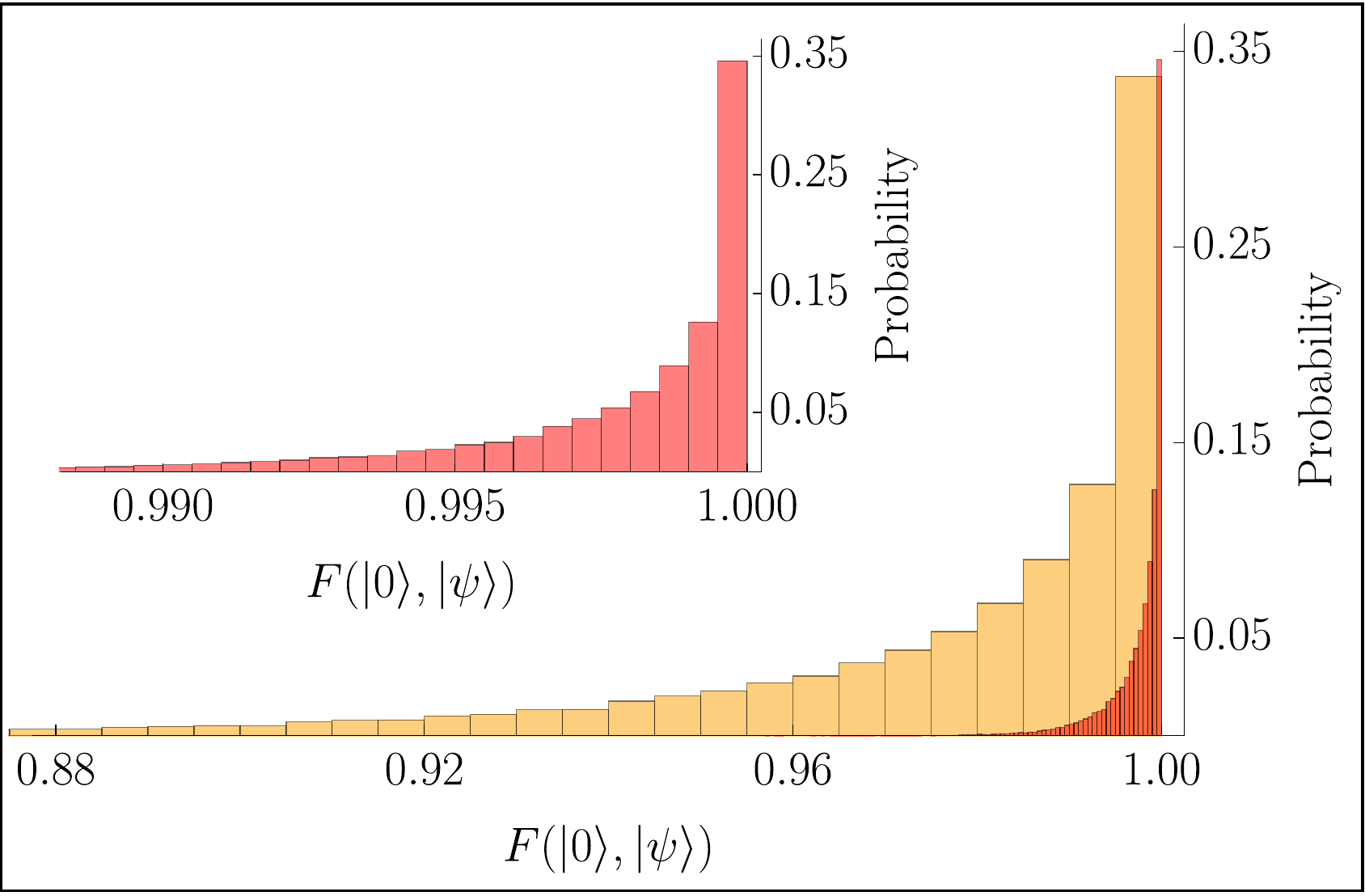}
\caption{\textbf{Visualization of von Mises-Fisher distributions}. Top: 50000 random states selected according to von Mises-Fisher distribution around $\ket{0}$ state (red point) with left: $\kappa = 10$ and right: $\kappa=100$. Bottom: Fidelity between $\ket{0}$ state and states $\ket{\psi}$ that are drawn from  von Mises-Fisher distribution (i.e. between the ``red point'' state and states represented by ``blue points'') expressed in the form of histograms (orange: $\kappa=10$, red (and inset): $\kappa=100$).}\label{fig:bloch_vonMises}
\end{figure}

In directional statistics \cite{mardia2009directional} von Mises-Fisher distribution \cite{fisherDispersionSphere1953} is a continuous probability distribution on the $N$-dimensional sphere (see Fig.~\ref{fig:bloch_vonMises}), and plays a similar role to  a normal distribution on a flat manifolds. Since pure states of qubits live on a Bloch sphere, it is more natural to exploit directional statistics and use von Mises-Fisher distribution as a distribution for initial state preparation than standard normal distribution.  For a 2-sphere the probability density function is given by  
\begin{equation}\label{eq:mises}
    p(\vec{x},\vec{\mu},\kappa) = \frac{\kappa}{4\pi \sinh \kappa} e^{\kappa \vec{\mu} \cdot \vec{x}},
\end{equation}
where $\vec{\mu}, \vec{x}$ are normalized vectors, and $\kappa$ is similar to inverse of the variance - for $\kappa\to 0$ it converges to a uniform distribution, while for $\kappa\to +\infty$ it is localized around $\vec{\mu}$, which resembles mean value in the standard normal distribution. Note that Eq.~(\ref{eq:mises})    is a special case of Kent distribution \cite{KentFisher-BinghamDistributionOnSphere}.

If we fix $\vec{\mu}=(0,0,1)$ (see Fig.~\ref{fig:bloch_vonMises}), which corresponds to the distribution around the North Pole, i.e. around the $\ket{0}$ state,  
the averaging of the fidelity is 
with respect to the following normalized surface element $d \Omega$ 
\begin{equation}
\int d\Omega = \int_{\theta=0}^\pi \int_{\phi=0}^{2\pi}\frac{\kappa \sin\theta}{4\pi\sinh\kappa}e^{\kappa \cos\theta}d\theta d\phi = 1.
\end{equation}
Integrating $\tr\big[U\ket{\psi}\bra{\psi}U^\dag\mathcal{E}_U(\ket{\psi}\bra{\psi})\big]$ with respect to von Mises-Fisher surface element one arrives at
\begin{eqnarray}
\bar{F}_{\kappa} &=& \frac{1}{2}+ \frac{\kappa \coth \kappa - 1}{4\kappa^2}\sum_{k=1}^2\tr \Big(U \sigma_kU^\dag \mathcal{E}_U \big[\sigma_k\big]\Big) +\nonumber\\
 &+&\frac{2-2\kappa \coth \kappa + \kappa^2}{4\kappa^2}\tr \Big(U \sigma_3U^\dag \mathcal{E}_U \big[\sigma_3\big]\Big) +\nonumber\\
 &+& \frac{\kappa \coth \kappa -1}{4\kappa}\tr \Big(U \sigma_3U^\dag \mathcal{E}_U \big[\sigma_0\big]\Big)\label{eq:fid_kappa},
\end{eqnarray}
which depends on the same $\chi$ matrix elements as in the polar cap case, i.e. contribution as in Eqs.~(\ref{eq:chi_el1}-\ref{eq:chi_el3}). Moreover, the same reasoning holds for distributions around different central states (e.g. $\ket{+})$.

\section{Results}
Now we demonstrate through our analytic expressions how certain noise channels that are indistinguishable under the standard average fidelity, Eq.~\eqref{eq:avg_fid}, can be distinguished via augmented fidelities, Eqs.~\eqref{eq:fid_theta},~\eqref{eq:fid_kappa}. Additionally, we analytically derive the spread between the maximal and minimal value of fidelity attainable for these processes, and obtain error bars based on the standard deviation.

The uniform-average fidelity Eq.~(\ref{eq:chi00}) depends purely on a single element of the $\chi$ (process) matrix of the composed noisy process $\mathcal{E}=\mathcal{U}^\dag\circ \mathcal{E}_U$.
Since the two augmented fidelities, Eqs.~\eqref{eq:fid_theta},~\eqref{eq:fid_kappa}, are influenced by $\chi$ matrix elements present in Eqs.~(\ref{eq:chi_el1}-\ref{eq:chi_el3}) (for distributions centered around $\ket{0}$) it suffices to consider a matrix of the following form
\begin{equation}\label{eq:chi_restricted}
    \chi = \left(\begin{array}{cccc}\chi_{00} & \cdot & \cdot & \chi_{03} \\\cdot & \chi_{11} & -i \chi_{03} & \cdot \\\cdot & i \chi_{03} & \chi_{22} & \cdot \\ \chi_{03} & \cdot & \cdot & \chi_{33}\end{array}\right),
\end{equation}
where the $\chi_{ij}$ elements are real by hermiticity. 
The dots in Eq.~\eqref{eq:chi_restricted} indicate these elements are arbitrary for our purposes, as they are absent in Eqs.~\eqref{eq:fid_theta} and \eqref{eq:fid_kappa} (up to $\chi$ being a genuine process matrix). For simplicity, we set these elements to zero. With this, constraints to impose CPTP conditions are $\sum_k \chi_{kk}=1$, and $\chi\ge0$ which translates into $\chi_{00}\chi_{33}\ge \chi_{03}^2$ and $\chi_{11}\chi_{22}\ge \chi_{03}^2$. 

\begin{figure}[t]
\includegraphics[width=0.4\textwidth]{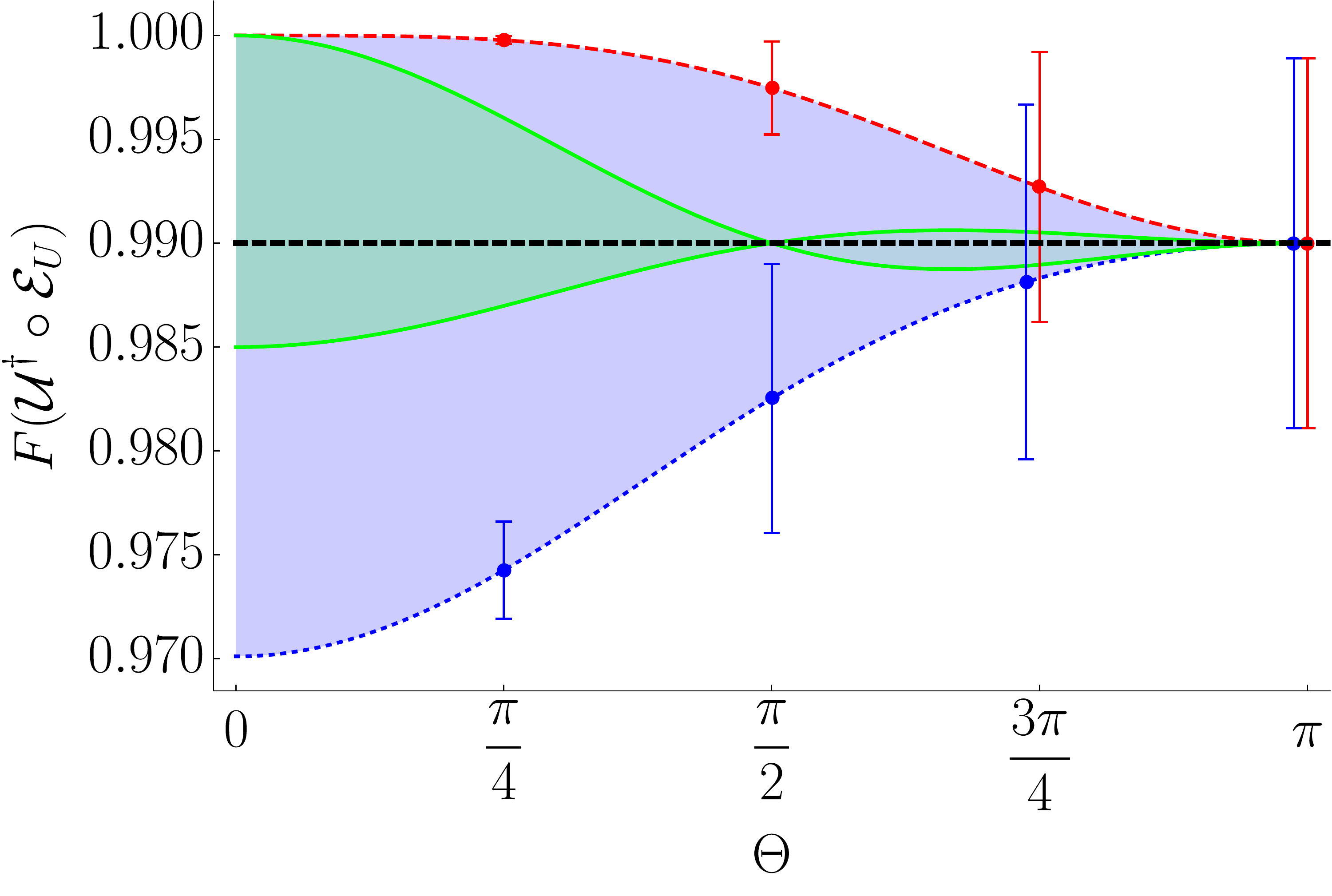}
\includegraphics[width=0.4\textwidth]{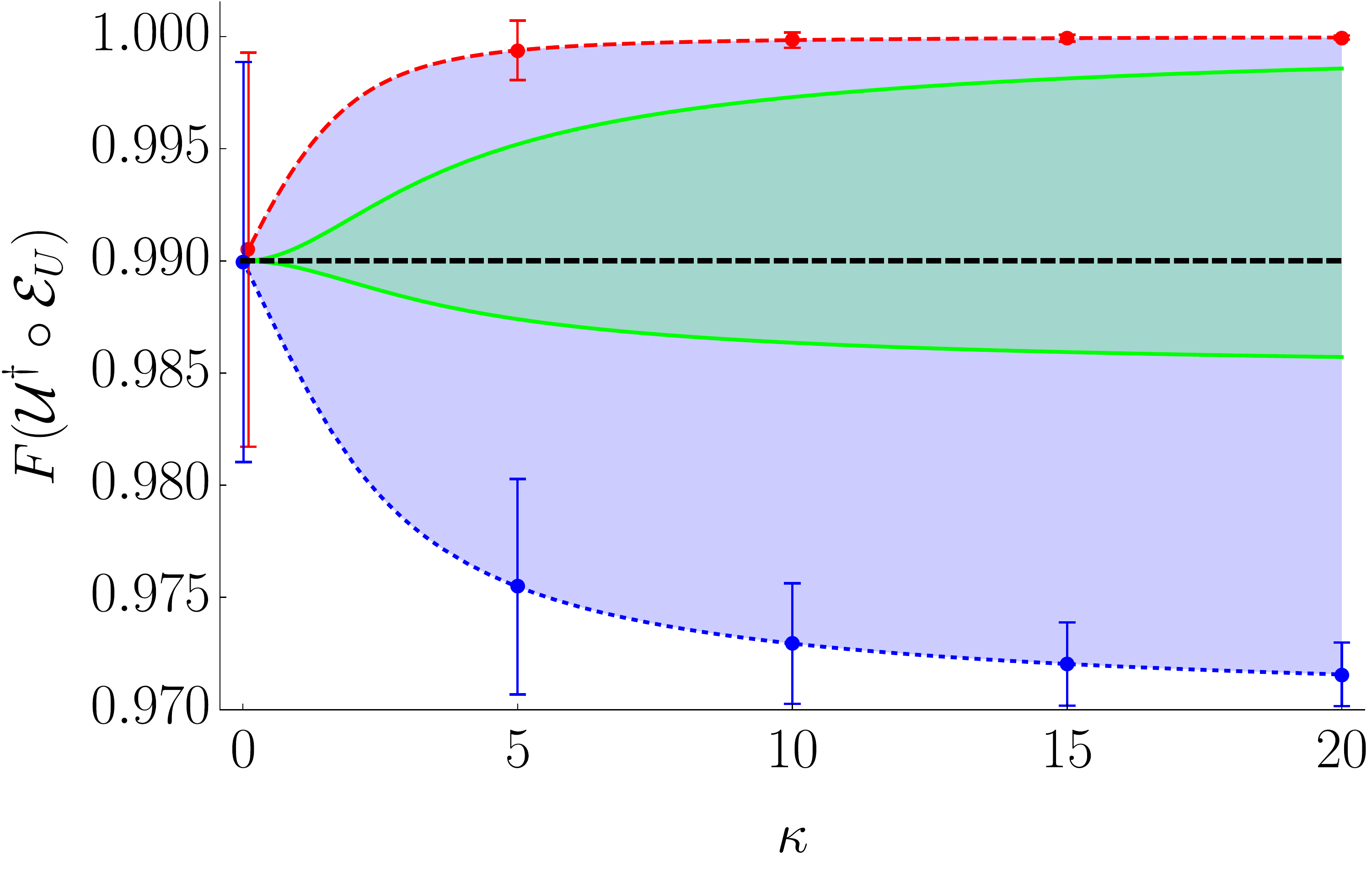}
\caption{{\bf Analytical 
augmented fidelities for processes that share the same $\mathbf{\chi_{00}=0.985}$ element (depolarizing rate) corresponding to 99\% fidelity}. 
Red (dashed) and blue (dotted) curves  bound the region between the minimal and maximal fidelities of a noisy process. Green region (inside) corresponds to diagonal $\chi$ matrices (i.e. Pauli channels).
Top: augmented fidelity over polar cap distribution, Eq.~\eqref{eq:fid_theta}, parametrized by polar angle $\Theta$, bottom: 
augmented fidelity,  Eq.~\eqref{eq:fid_kappa}, with respect to von Mises-Fisher distribution (normal distribution in directional statistics) as a function of $\kappa$ parameters (corresponding to the inverse of variance in standard statistics for normal distribution). Error bars indicate standard deviation of fidelities (color coded) for minimal and maximal values. Black (dashed) line represents the depolarizing channel that would be identified through RB.}\label{fig:fid_theta_kappa}
\end{figure}

In order to investigate the spread of fidelities, we need to minimize (maximize) the average fidelity, according to Eqs.~(\ref{eq:fid_theta}) and (\ref{eq:fid_kappa}). First let us introduce $p=(1-\sqrt{\chi_{00}})^2$. We can analytically determine the minimal (maximal) value, which is achieved for $\chi_{33}=p,\ \chi_{11}=\chi_{22}=\sqrt{p}-p$ and $\chi_{03} = -\chi_{11}$ (respectively $\chi_{03}=\chi_{11}$). 
A similar analysis can be performed for a Pauli channel, i.e. with diagonal $\chi$ matrix. In that case, the minimal fidelity values are achieved for $\chi_{11}=1-\chi_{00}$ (or $\chi_{22}=1-\chi_{00})$ and maximal for $\chi_{33}=1-\chi_{00}$.

The spread between minimal and maximal fidelities for noise models with the uniform-average fidelity of 99\% (corresponding to $\chi_{00}=0.985$) is depicted in Fig.~\ref{fig:fid_theta_kappa}. In the limit of $\kappa\to0$ and $\Theta=\pi$ we recover results for average over all states. Note, that in that case spread completely disappears and it is impossible to differentiate between various noise models (i.e. sharing the same $\chi_{00}$ element, but otherwise having distinct elements of the $\chi$ matrix) solely based on fidelity Eq.~(\ref{eq:avg_fid}). On the other end, when $\kappa\to\infty$ and $\Theta\to0$ both distributions tend to a localized state $\ket{0}$ and fidelities display the largest spread. Moreover, the spread increases with decreasing $\chi_{00}$, i.e. higher uniform-average infidelity allows for larger spread. Therefore this analysis provides a trade-off between faithful state preparation (more localized distribution) and sensitivity to noise manifested by fidelities. 

As we have emphasized, while 
standard benchmarking techniques (such as randomized benchmarking (RB) \cite{emersonScalableNoiseEstimation2005}) 
probe the uniform-average fidelity, identifying the $\chi_{00}$ element, but do not detect properties of noise presence encoded elsewhere in the $\chi$ matrix. 
Additionally, it is well known that a twirling protocol (see for example \cite{emersonScalableNoiseEstimation2005,moussaPracticalExperimentalCertification2012} and references therein), which is a mathematical justification for RB methods, is insensitive to initial state distribution; since it averages over the entire unitary group, it transforms each channel into a depolarizing one. Thus, after performing a twirling protocol, one is left with a contribution stemming only from the $\chi_{00}$ element (i.e. depolarizing rate).
It is attractive, yet incorrect, to equate every noise process with depolarizing noise, simplifying the entire noise analysis to this averaged case.

For each distribution one can also determine variance $\sigma^2(F)$ of the fidelity
\begin{equation}
\label{eq:variance}
    \sigma^2(F) = \int F_{|\psi\rangle\langle\psi|}(U,\mathcal{E}_U)^2 d\Omega-\Big(\int  F_{|\psi\rangle\langle\psi|}(U,\mathcal{E}_U) d\Omega\Big)^2, 
\end{equation}
where $F_{|\psi\rangle\langle\psi|}(U,\mathcal{E}_U)$ is given by Eq.~(\ref{eq:single_state_fidelity})
and $d\Omega$ is a surface element related to the underlying distribution of initial states $\ket{\psi}$.
In \cite{magesanGateFidelityFluctuations2011} the formula for the uniform distribution over the entire space was provided. 
It is also straightforward to calculate this in the case of polar cap and von Mises-Fisher distributions (see Appendix), which depends not only on the $\theta$ and $\kappa$ parameters, but in general on all $\chi$ matrix elements. The variance becomes smaller for more localized distributions (i.e. as $\kappa\to\infty$ and $\Theta\to0$) and reaches its largest value for distributions close to uniform over all states.

In Fig.~\ref{fig:fid_theta_kappa} we report standard deviation error bars (i.e. from Eq.~\eqref{eq:variance}) for the minimal (blue) and maximal (red) channels. This corresponds to the spread in fidelity values over the pure states of the distributions, for these extreme channels, and depends on all elements of Eq.~\eqref{eq:chi_restricted}.

Our analysis shows that we have two independent sources of fidelity deviations. One related to a statistical distribution, and the second one to noise process. In principle they can either benefit (increase fidelity) or hamper (decrease) the performance. It is important to properly identify their impact and origins. 

\subsection{Noise bias}
\begin{figure}[!htb]
\includegraphics[width=0.4\textwidth]{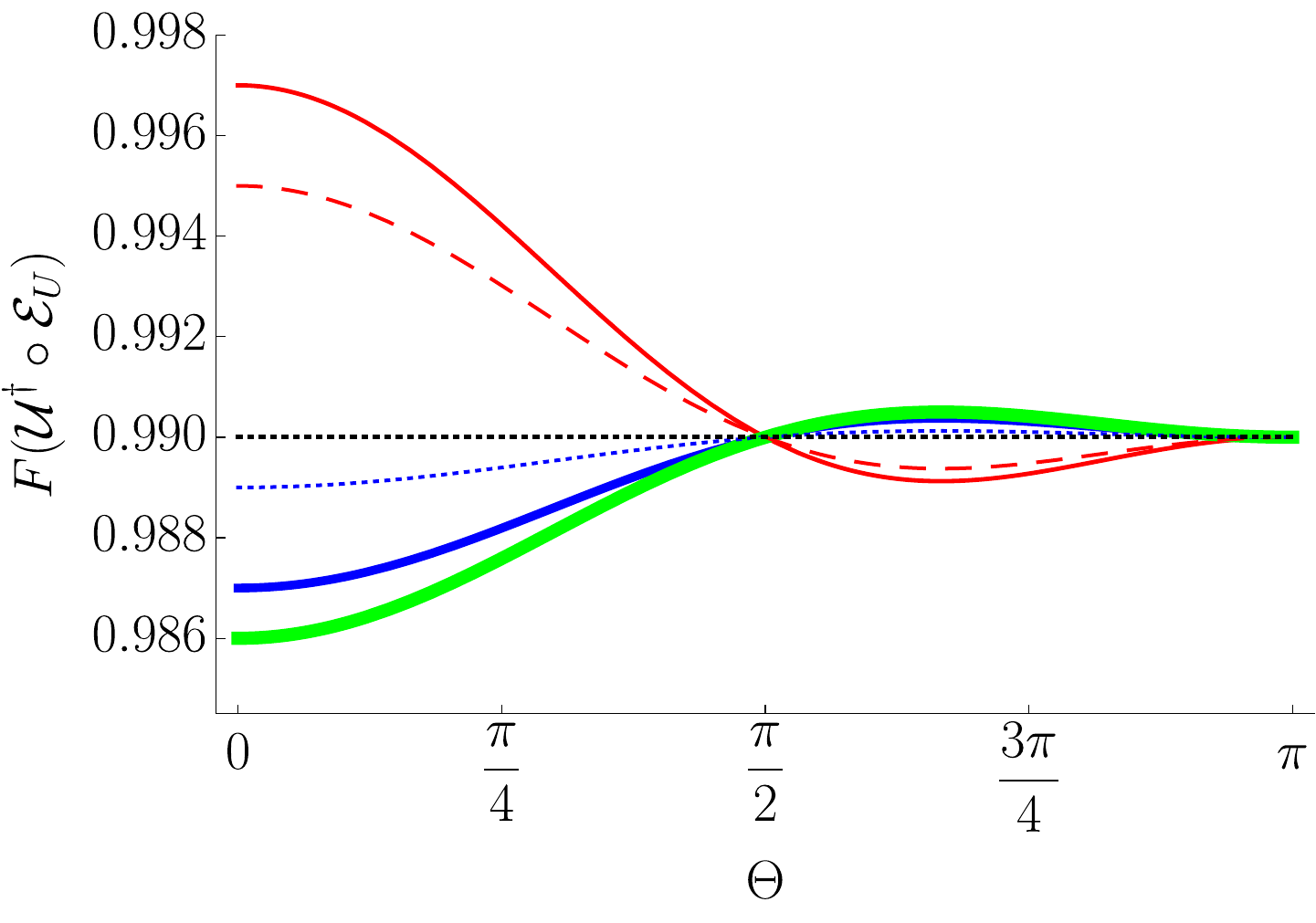}
\includegraphics[width=0.4\textwidth]{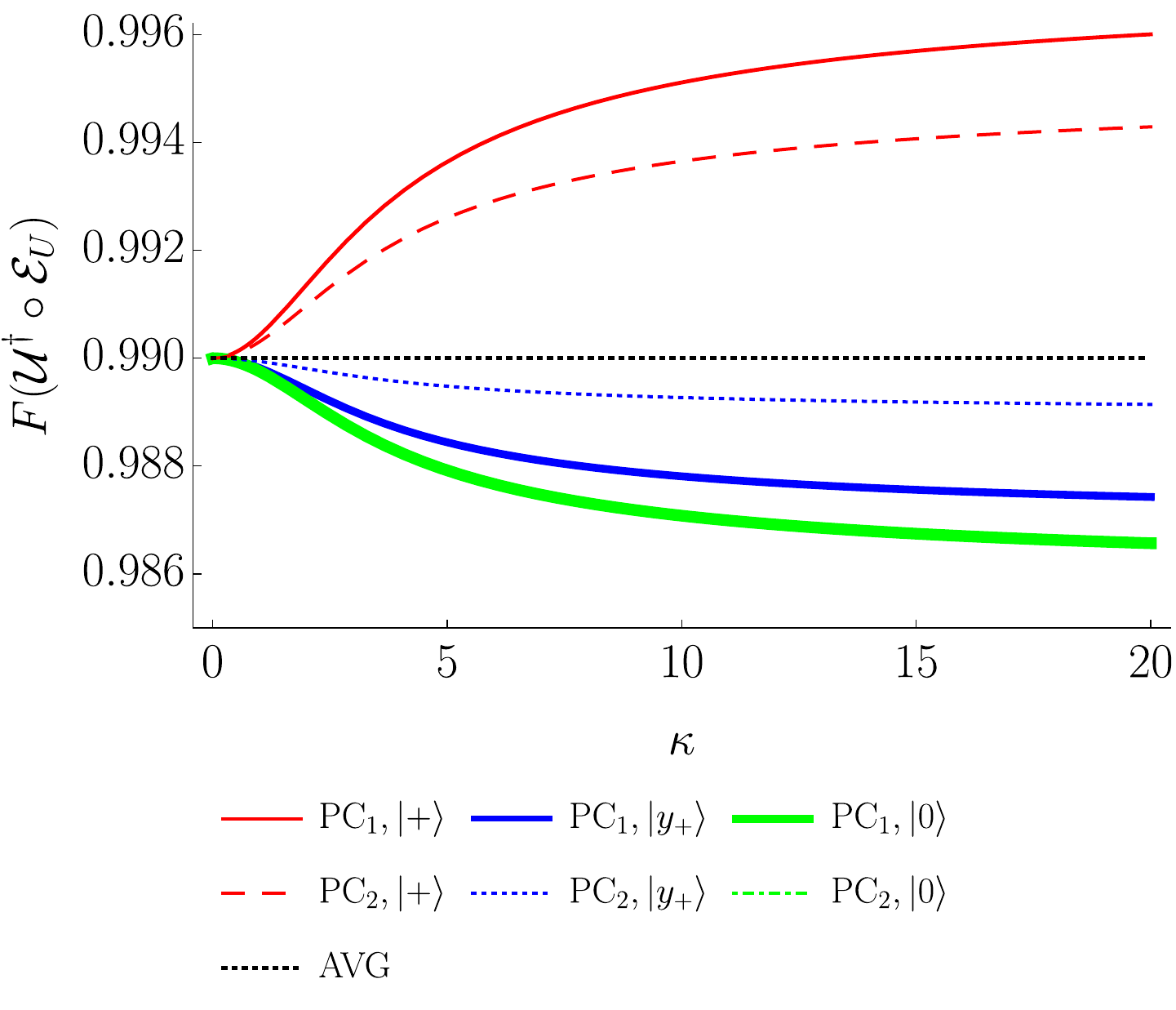}
\caption{\textbf{Augmented fidelity of Pauli channels.} Different channels (line styles)  characterized by values in Table~\ref{tab:noise_bias} with polar cap (top panel) and von Mises-Fisher (bottom) distributions. Black dashed line corresponds to the depolarizing channel with fidelity 99\%. Color coded are distribution centered around different initial states: (red) $\ket{+}$ state, (blue) $\ket{y_+}$ (eigenvector of Pauli $Y$ to eigenvalue +1, and (green) $\ket{0}$. 
}\label{fig:noise_bias}
\end{figure}

Recently the problem of noise bias in Pauli channels, i.e. having diagonal $\chi$ matrix, has attracted considerable attention, especially in the field of error correction \cite{aliferis2009fault,TuckettPRLUltrahigh2018, TuckettPRXTailoring2019}. The noise bias (in $Z$ direction) is defined for Pauli channels, as $\eta_Z = \chi_{33}/(\chi_{11}+\chi_{22})$, and informs us which Pauli error is more prominent. Here, we propose to identify Pauli errors by looking at the average gate fidelity either with polar cap or von Mises-Fisher distribution centered around eigenstates of $X,Y$ and $Z$ to the eigenvalue $+1$. In Table~\ref{tab:noise_bias} we report values of $\chi_{kk}$ elements for two different Pauli channels (which we call PC$_1$ and PC$_2$) that take the same value of $\chi_{00}=0.985$, i.e. corresponding to uniform-average fidelity of $99\%$, and the same noise bias $\eta_Z=1/14$.
\begin{table}[!htb]
\setlength{\tabcolsep}{14pt}
\caption{}\label{tab:noise_bias}
\begin{tabular}{@{}*{5}{c}}\toprule

ID   & $\chi_{11}$ & $\chi_{22}$ & $\chi_{33}$ & $\eta_Z$ \\ \midrule\addlinespace
PC$_1$ & 0.012 & 0.002 & 0.001 & $\frac{1}{14}$ \\\noalign{\smallskip}  
PC$_2$  & 0.010 & 0.004 & 0.001 & $\frac{1}{14}$ \\\noalign{\smallskip}
\bottomrule
\end{tabular}
\end{table}
The results for these channels are displayed in  Fig.~\ref{fig:noise_bias}.
Note, that for distributions centered around $\ket{0}$ (green color  in Fig.~\ref{fig:noise_bias}) we see only a single line style. This is due to the fact that augmented fidelity fails to discriminate   $Z$ bias in this protocol. However, if one changes the center of the distribution, the difference becomes 
clear and two Pauli channels yield distinct values. Note that changing the center is due to a special type of single-qubit rotation transformation, and therefore could be also used as a benchmark for single qubit rotation gates (the profile of the curves are qualitatively the same as in Fig.~\ref{fig:noise_bias}).  

\subsection{2-qubit Case \label{sect:2q}}
We consider special class of two-qubit fidelities with 
\begin{equation}
\begin{split}
    &F(U,\mathcal{E}_U) = \\ &\Tr\Big(U\ket{\psi_1}\bra{\psi_1}\otimes \ket{\psi_2}\bra{\psi_2} U^\dag \mathcal{E}_U(\ket{\psi_1}\bra{\psi_1}\otimes \ket{\psi_2}\bra{\psi_2})\Big),
\end{split}
\end{equation}
and now taking the average only over the local distributions
\begin{equation}\label{eq:2q_total_fid}
    \bar{F}(U,\mathcal{E}_U) = \int F(U,\mathcal{E}_U)d\Omega_1 d\Omega_2,
\end{equation}
where $d\Omega_k$ corresponds to the surface element associated with qubit $k$. If both $\ket{\psi_1}$ and $\ket{\psi_2}$ are uniformly distributed over the entire space, then Eq.~\eqref{eq:2q_total_fid} yields
\begin{equation}\label{eq:2q_uniform}
\begin{split}
     \bar{F} = \frac{1}{9}\Big(1+8\chi_{00,00}+2(\chi_{01,01}+\chi_{02,02}+\chi_{03,03}+ \\  \chi_{10,10}+\chi_{20,20}+\chi_{30,30})\Big),
\end{split}
\end{equation}

where we use map in the form
\begin{equation}
    \mathcal{U}^\dag\circ\mathcal{E}_U(\rho) = \sum_{k,l,m,n=0}^3 \chi_{kl,mn}\big(\sigma_k\otimes\sigma_l\big)\rho\big( \sigma_m\otimes\sigma_n\big).
\end{equation}
Note that Eq.~\eqref{eq:2q_uniform} depends not only on the $\chi_{00,00}$ element as in the case of full space uniform distribution, but also on the other (diagonal) elements of the $\chi$ matrix. 

\begin{figure}[t!]
\begin{tabular}{cc}
\includegraphics[width=42mm]{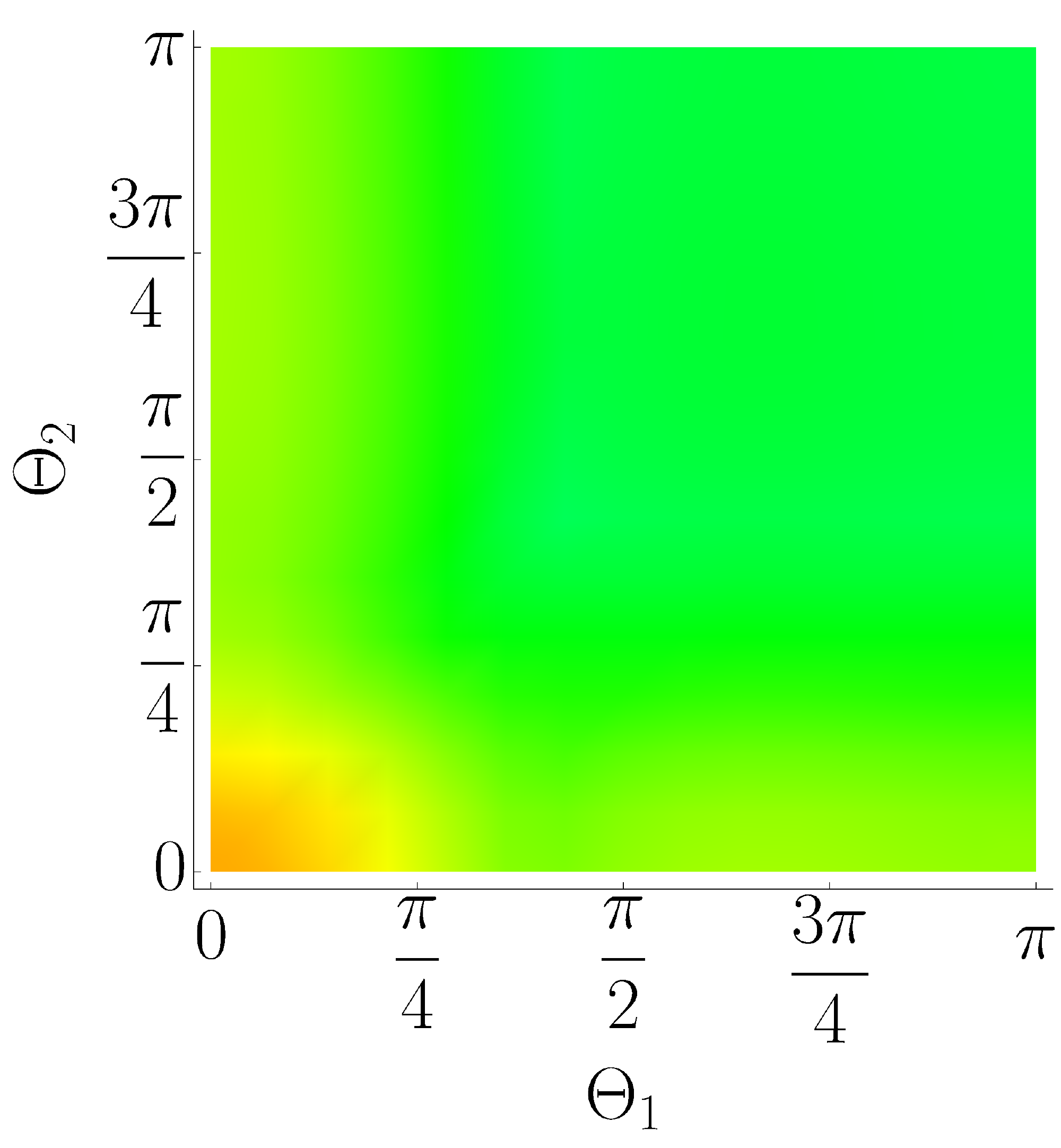} & \includegraphics[width=42mm]{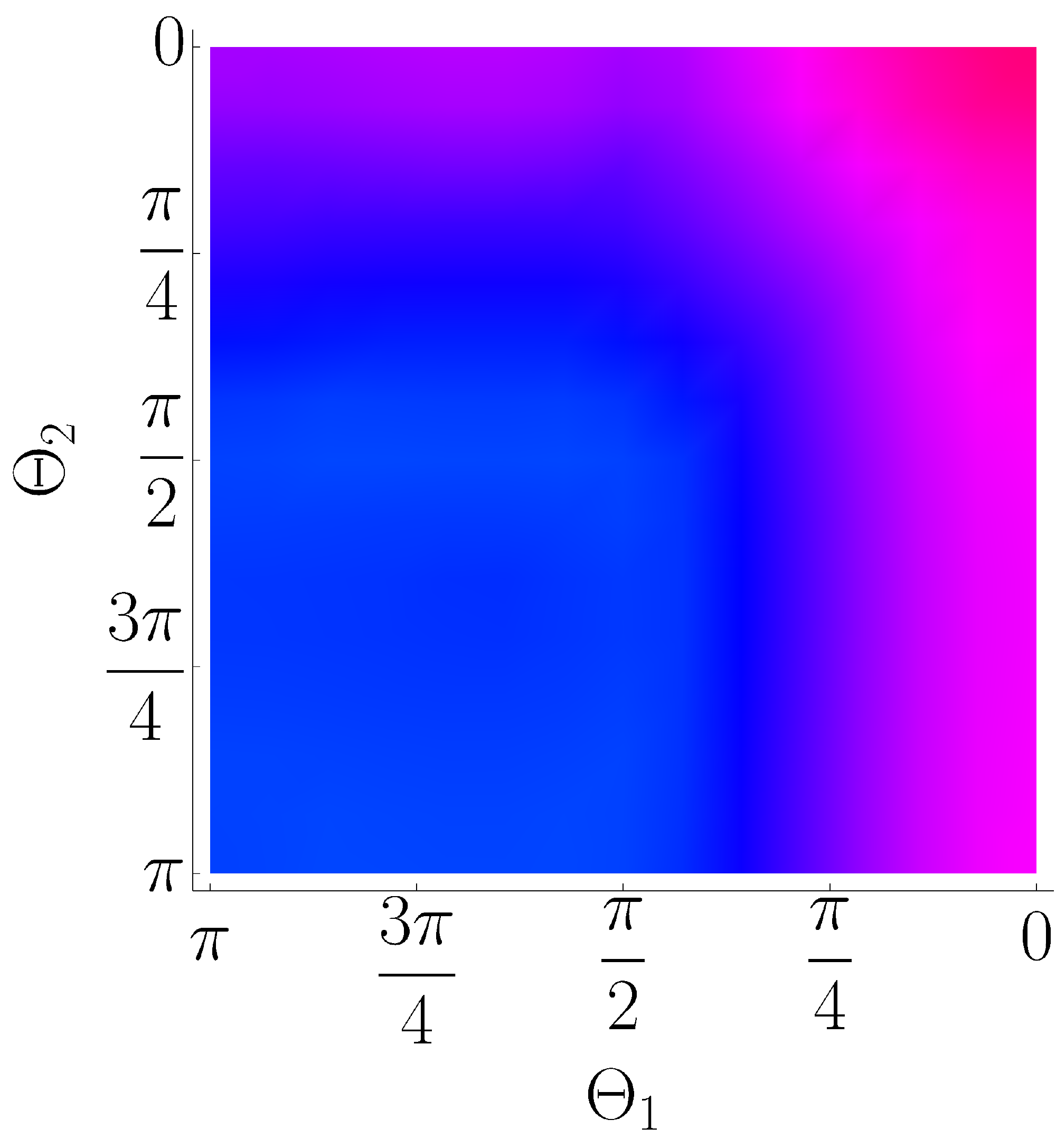}\\
\includegraphics[width=42mm]{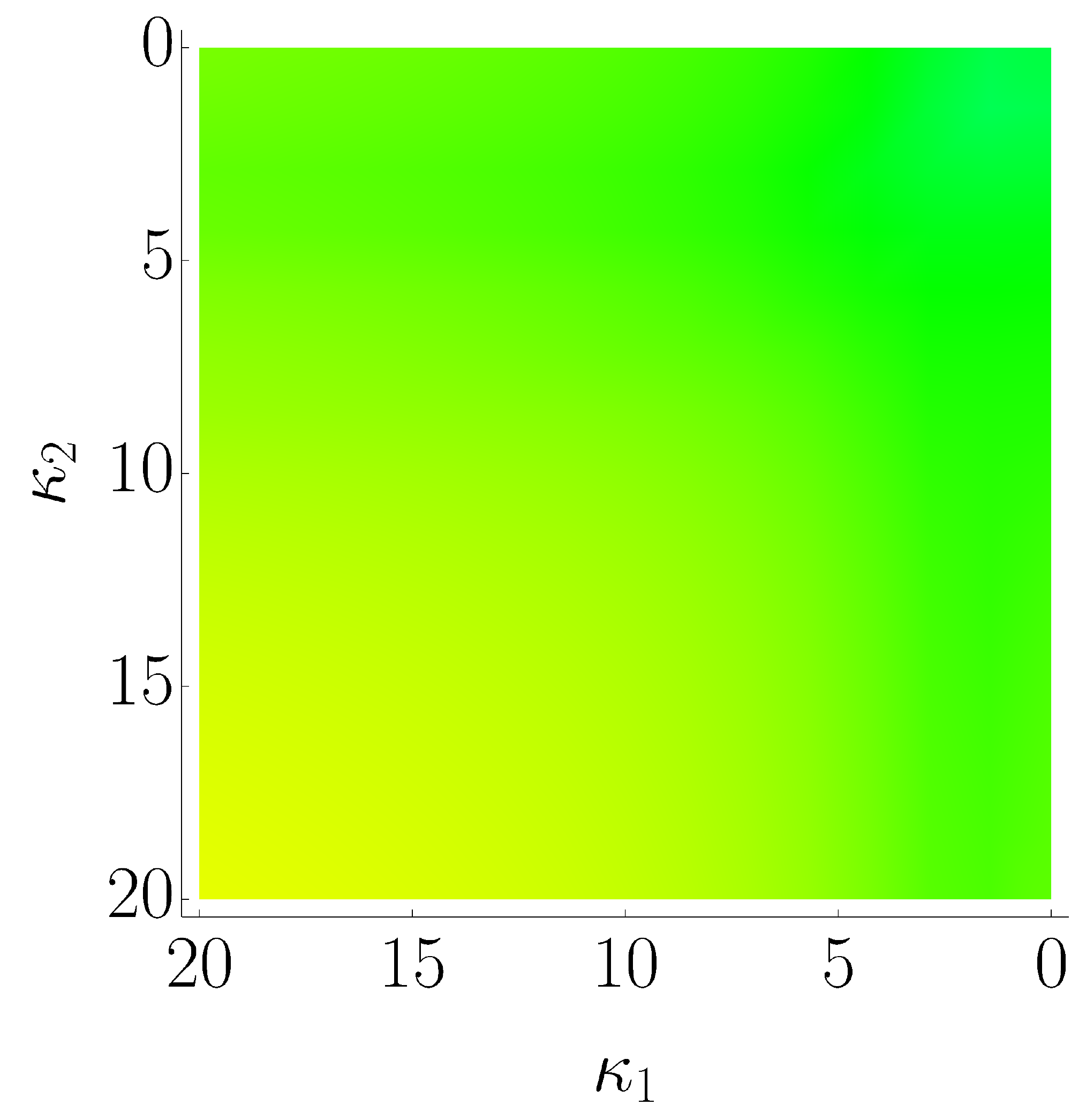} & \includegraphics[width=42mm]{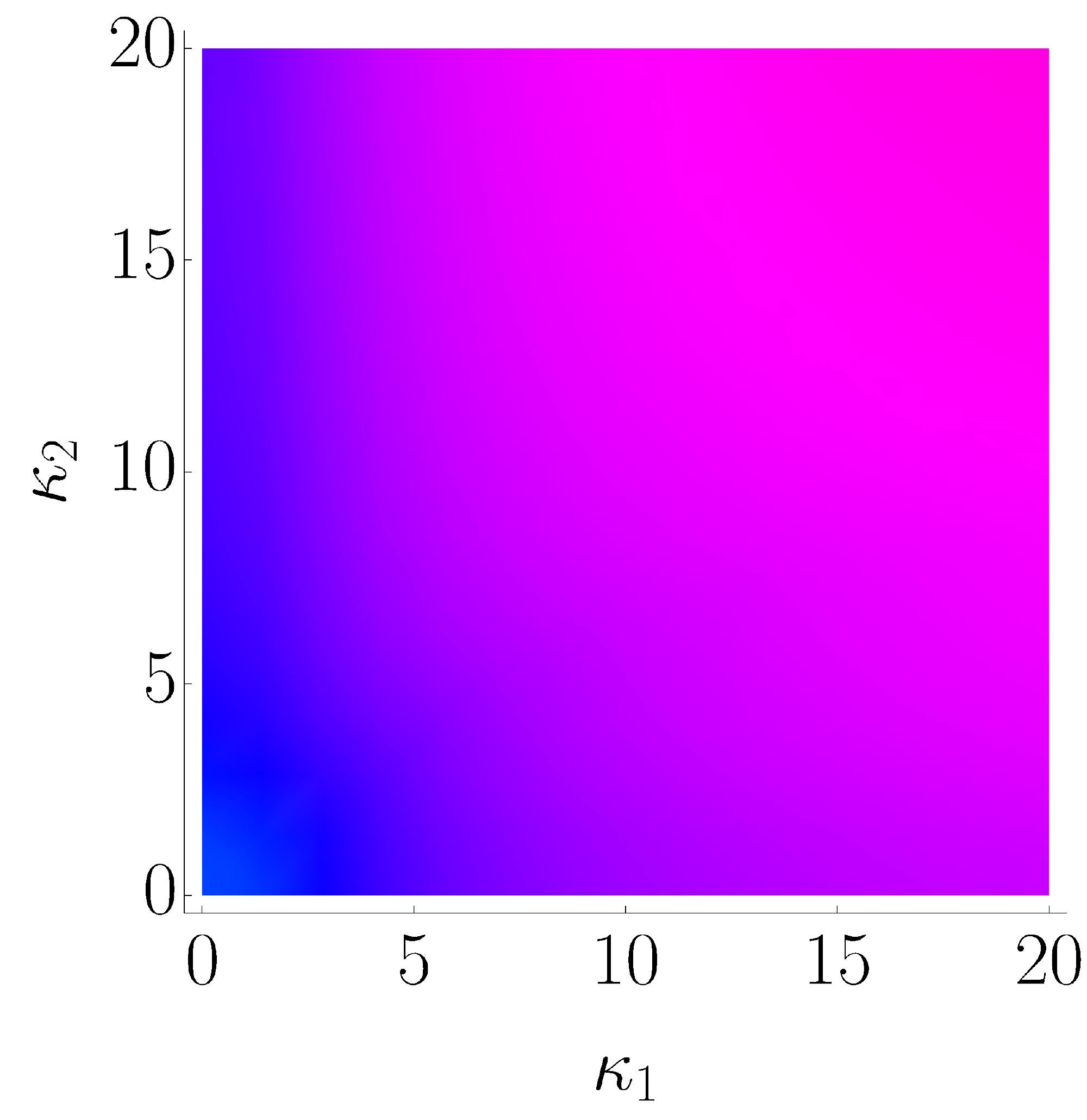}\\
\multicolumn{2}{c}{\includegraphics[width=80mm]{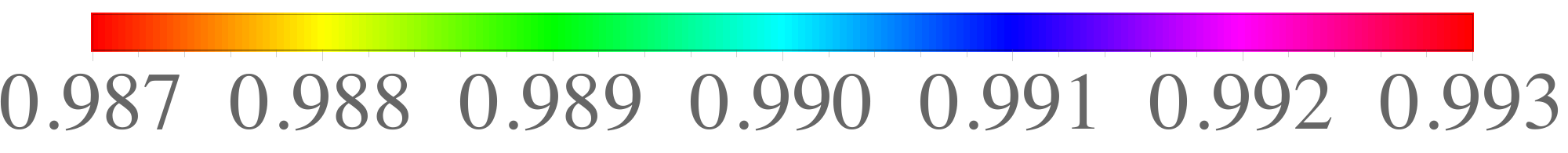} }
\end{tabular}
    \caption{\textbf{Heatmaps of two-qubit augmented fidelities for local polar cap distribution (top panel) and von Mises-Fisher distribution (bottom)}.
    The left/right panel(s) show the minimum/maximum augmented fidelity envelope over with all process matrices considered. 
    Note, that in the limit $\Theta_{1,2}\to\pi$ and $\kappa_{1,2}\to 0$, the augmented fidelity fails to reach 99\% fidelity, which is related to Eq.~\eqref{eq:2q_uniform} and the additional contribution of other diagonal elements.}\label{fig:heatmaps}

\end{figure}


For local distributions of polar cap and von Mises-Fisher type we perform numerical analysis based on 1673 random process matrices \footnote{A random $\chi$ matrix was generated by fixing the $\chi_{00,00}$ element, drawing randomly a vector of positive numbers that sum up to 0.015 as diagonal elements, and then drawing randomly off-diagonal elements between [-min\{diagonal elements\}, min\{diagonal elements\} and rejecting samples when $\chi<0$. From total 6000 samples, we accepted 1673 positive definite matrices that yield CPTP condition of the map.} with the same unitarily invariant element $\chi_{00,00}=0.985$ (corresponding to 99\% unifrom-average fidelity). The results are displayed in Fig.~\ref{fig:heatmaps}, which show variations in the augmented fidelities through changing the distribution parameters (here two angles $\Theta_{1,2}$, or von Mises-Fisher inverse variance $\kappa_{1,2}$).

\section{Conclusions}
In this article we examined and calculated augmented fidelities of noisy single qubit gates by averaging over different initial state distributions. In particular we focused on two models 
-- polar cap and von Mises-Fisher distributions, parametrized by a polar angle $\Theta$ and variance-related parameter $\kappa$, respectively. The introduced methods augment the uniform-average fidelity (strongly based on Haar invariance property), and carry additional information about underlying noise process. This information is manifested in larger possible spread in observed fidelities, that also allows to identify noise biases in Pauli Channels. Because the uniform-average fidelity and associated protocols 
probe only the depolarizing character of the noise, 
they can under or overestimate the deteriorating effect of device's miscalibration.  
Therefore, it is imperative to have 
additional tools for assessing performance of quantum devices. In particular, efficient methods that reliably infer more $\chi$ matrix elements could improve functionality of next generations of quantum hardware.

Since it is impossible to perfectly prepare an arbitrary pure state, this method may also lend itself to probing state preparation errors.

Any reasonably effective pure state preparation will likely resemble some possibly skewed distribution centered around the target state. This means that computing the fidelity for a particular state is similar to using our non-uniform distribution fidelity metrics. Note that the Kent distribution (which generalizes von Mises-Fisher) could in principle also be used to achieve a greater agreement between theory and experiment, especially in case when the underlying distribution displays anisotropic properties.  

As demonstrated in Figs.~\ref{fig:fid_theta_kappa}, \ref{fig:noise_bias}, the greatest spread and smallest error bars in our augmented fidelity metrics between noise channels occurs for a point distribution (i.e. at $\Theta=0=\kappa^{-1}$). As discussed in the previous paragraph, this is not a realistic scenario to probe, and this effectively sets a lower bound on $\Theta$ and $\kappa^{-1}$ based on experimental capabilities. As $\Theta$ and $\kappa^{-1}$ are increased however, the ability to discriminate channels decreases. Therefore it is important to be able to determine in experiment what reasonable lower bounds are on these quantities.

Lastly we mention that the introduced figures of merit can be measured experimentally with current technology as a slight modification to current techniques.
Indeed, similar to the uniform-average fidelity (see \cite{bowdreyFidelitySingleQubit2002}), one may restrict to performing state tomography along 6 initial states, i.e. in the $\pm x,\pm y, \pm z$ directions on the Bloch sphere. This would of course mean that state preparation and measurement errors are included in $\mathcal{E}_U$, though certain techniques may allow one to mitigate these effects (see for example \cite{spam-PT, sun2018efficient}). However, the number of measurements required in this case is larger than the number of measurements for process tomography (6 (states) $\times$ 3 (measurements per state) =18 for the first method and 12 for QPT).
Therefore, we leave construction of a more efficient protocol in the single qubit and higher dimensional cases to future work.

One promising research direction is to explore metrics corresponding non-uniform distributions in higher dimensions, investigating their capabilities to discriminate different noise channels. In this work, we restricted to distribution defined on the product states, where even local uniform distribution can provide additional insight about the noise process. Additionally, we explored (numerically) von Mises-Fisher and polar cap distributions for product states, demonstrating that conjugation of these techniques with other benchmarking methods can improve our understanding of the device imperfections. The generalization to nonlocal case is not straightforward, and  would need to take into account non-trivial geometrical structure of higher-dimensional pure quantum states \cite{bengtsson2017geometry}. The goal would be an efficient (in number of measurements) and scalable (in number of qubits) protocol that can reliably probe the proposed metrics. An auspicious direction is to examine protocols inspired by RB technique for local distributions, which would be sampled independently, hence offering scaling similar to single-qubit case. This we leave as an open problem for future research.  
Another aspect worth exploring is to use these types of distribution over a subspace of the full Hilbert space, such as is done in Refs.~\cite{pedersen2007fidelity,pedersen2008distribution} for the standard uniform-average fidelity metric. This is of particular interest as it is known higher system levels can play a dominant role in the projected two-level dynamics of a qubit system \cite{supremacy-blueprint}, and could in principle help us to identify leakage errors with better accuracy. Therefore having methods to better distinguish noise processes acting on the full $d$-level (qudit) system could have immediate implications for hardware design.

\acknowledgments
We appreciate fruitful discussion with Robin Blume-Kohout.  
We are grateful for support from NASA Ames Research Center, the AFRL Information Directorate under grant F4HBKC4162G001. 
FW and JM are thankful for support from NASA Academic Mission Services, Contract No. NNA16BD14C. AP is grateful to his QuAIL colleagues for kind hospitality during his work at NASA.

\appendix*
\section{Variance}
The variance of a fidelity is computed with respect to its distribution as
\begin{equation}
    Var(F) = \int F_{\ket{\psi}\bra{\psi}}(U,\mathcal{E}_U)^2 d\Omega-\Big(\int  F_{\ket{\psi}\bra{\psi}}(U,\mathcal{E}_U) d\Omega\Big)^2, 
\end{equation}
where 
\begin{equation}
    F_{\ket{\psi}\bra{\psi}}(U,\mathcal{E}_U) = \tr\Big[U \ket{\psi}\bra{\psi}U^\dag\mathcal{E}_U(\ket{\psi}\bra{\psi} \Big],
\end{equation}
and $d\Omega$ is a surface element related to the underlying distribution of initial states $\ket{\psi}$. Taking an arbitrary representation of a noise process (i.e. characterized by a $\chi$ matrix, see Eq.~(\ref{eq:map})) results in a formula that depends non-trivially on all elements of $\chi$ matrix. For our purposes it suffices to consider noise processes that minimize (maximize) fidelities, which means of the form Eq.~(\ref{eq:chi_restricted}). In that case   polar cap distribution variance for minimal fidelity is
\begin{widetext}
\begin{eqnarray*}
Var(F_\Theta) & = &\frac{1}{5760}\Big[-40 \Big(-12 \chi _{0,3} (\cos (\Theta )+1)+\big(\chi _{1,1}+\chi _{2,2}-2 \chi _{3,3}\big) (2 \cos (\Theta )+\cos (2 \Theta ))-6 \chi _{0,0}+3 \big(\chi _{1,1}+\chi _{2,2}-2\big)\Big){}^2 \\
&-& \frac{3}{\cos (\Theta )-1}\Big(-1920 \chi _{0,1} \chi _{1,3} \sin ^4(\Theta )+1920 \chi _{0,0}^2 \big(\cos (\Theta )-1\big)-120 \big(2 \chi _{0,2} \chi _{2,3}+\chi _{0,3} \left(\chi _{1,1}+\chi _{2,2}-2 \chi _{3,3}\right)\big) \cos (4 \Theta )\\
&+&3 \left(3 \chi
   _{1,1}^2+2 \chi _{2,2} \chi _{1,1}+3 \chi _{2,2}^2+8 \chi _{3,3}^2+4 \left(\chi _{1,2}^2-4 \left(\chi _{1,3}^2+\chi _{2,3}^2\right)\right)-8 \left(\chi _{1,1}+\chi _{2,2}\right) \chi _{3,3}\right) \cos (5 \Theta )\\
   &+&30 \big(96
   \chi _{0,2}^2+64 \chi _{0,3}^2+20 \chi _{1,2}^2+8 \chi _{3,3}^2+5 \left(3 \chi _{1,1}^2+2 \chi_{2,2} \chi _{1,1}+3 \chi _{2,2}^2\right)+16 \left(\chi _{1,3}^2+\chi _{2,3}^2\right)\\
   &+& 8 \left(\chi _{1,1}+\chi _{2,2}\right) \chi
   _{3,3}\big) \cos (\Theta )+5 \big(-64 \chi _{0,2}^2+128 \chi _{0,3}^2-20 \chi_{1,2}^2+24 \chi _{3,3}^2-5 \left(3 \chi _{1,1}^2+2 \chi _{2,2} \chi _{1,1}+3 \chi _{2,2}^2\right)\\
   &+& 16 \left(\chi _{1,3}^2+\chi _{2,3}^2\right)+8
   \left(\chi _{1,1}+\chi _{2,2}\right) \chi _{3,3}\big) \cos (3 \Theta )+480 \left(2 \chi _{0,2} \chi _{2,3}+\chi _{0,3} \left(\chi _{1,1}+\chi _{2,2}+2 \chi _{3,3}\right)\right) \cos (2 \Theta )\\
   &-&5120 \chi _{0,1}^2 \sin
   ^4\left(\frac{\Theta }{2}\right) (\cos (\Theta )+2)+160 \chi _{0,0} \big(-24 \chi _{0,3} \sin ^2(\Theta )+\left(\chi _{1,1}+\chi _{2,2}\right) (9 \cos (\Theta )-\cos (3 \Theta )-8)\\
   &+&2 \chi _{3,3} (3 \cos (\Theta )+\cos (3 \Theta
   )-4)\big)-384 \chi _{3,3}^2-8 \big(320 \chi _{0,2}^2+90 \chi _{2,3} \chi _{0,2}+320 \chi _{0,3}^2+45 \chi _{0,3} \left(\chi _{1,1}+\chi _{2,2}\right)\\
   &+&16 \left(3 \chi _{1,1}^2+2 \chi _{2,2} \chi _{1,1}+3 \chi _{2,2}^2+4
   \left(\chi _{1,2}^2+\chi _{1,3}^2+\chi _{2,3}^2\right)\right)\big)-16 \left(75 \chi _{0,3}+16 \left(\chi _{1,1}+\chi _{2,2}\right)\right) \chi _{3,3} \Big) \Big],
\end{eqnarray*}
\end{widetext}
and for von Mises-Fisher distribution
\begin{widetext}
\begin{eqnarray*}
Var(F_\kappa) & = & \frac{1}{4 \kappa ^4}\Big[3 \chi _{0,0}^2 \kappa ^4-\chi _{1,1}^2 \kappa ^4-\chi _{2,2}^2 \kappa ^4+3 \chi _{3,3}^2 \kappa ^4+2 \chi _{1,1} \kappa ^4-2 \chi _{1,1} \chi _{2,2} \kappa ^4+2 \chi _{2,2} \kappa ^4+2 \chi _{1,1} \chi _{3,3} \kappa ^4+2 \chi
   _{2,2} \chi _{3,3} \kappa ^4\\
   &-&2 \chi _{3,3} \kappa ^4-\kappa ^4+8 \chi _{0,3} \kappa ^3-8 \chi _{0,3} \chi _{1,1} \kappa ^3-8 \chi _{0,3} \chi _{2,2} \kappa ^3+32 \chi _{0,2} \chi _{2,3} \kappa ^3-8 \chi _{0,3} \chi _{3,3} \kappa
   ^3+16 (\kappa  \coth (\kappa )-1) \chi _{0,1}^2 \kappa ^2\\
   &-&16 \chi _{0,2}^2 \kappa ^2+16 \chi _{0,3}^2 \kappa ^2+4 \chi _{1,1}^2 \kappa ^2+16 \chi _{1,2}^2 \kappa ^2-80 \chi _{1,3}^2 \kappa ^2+4 \chi _{2,2}^2 \kappa ^2-80 \chi
   _{2,3}^2 \kappa ^2-4 \text{csch}^2(\kappa ) \\
   &\cdot&\left(2 \kappa  \chi _{0,3}+\chi _{1,1}+\chi _{2,2}-2 \chi _{3,3}\right){}^2 \kappa ^2+24 \chi _{3,3}^2 \kappa ^2+4 \chi _{1,1} \kappa ^2-8 \chi _{1,1} \chi _{2,2} \kappa ^2+4 \chi
   _{2,2} \kappa ^2-12 \chi _{1,1} \chi _{3,3} \kappa ^2\\
   &-&12 \chi _{2,2} \chi _{3,3} \kappa ^2-8 \chi _{3,3} \kappa ^2+2 \chi _{0,0} \big(-\kappa ^2+\big(4 (\kappa  \coth (\kappa )-1) \chi _{0,3}+(\kappa +2 \coth (\kappa ))
   \left(\chi _{1,1}+\chi _{2,2}\right)\\
   &+&(3 \kappa -4 \coth (\kappa )) \chi _{3,3}\big) \kappa -2 \chi _{1,1}-2 \chi _{2,2}+4 \chi _{3,3}\big) \kappa ^2+32 \chi _{0,3} \chi _{1,1} \kappa +32 \left(\kappa ^2-3 \coth (\kappa )
   \kappa +3\right) \chi _{0,1} \chi _{1,3} \kappa \\
   &+&32 \chi _{0,3} \chi _{2,2} \kappa +96 \chi _{0,2} \chi _{2,3} \kappa -64 \chi _{0,3} \chi _{3,3} \kappa +4 \coth (\kappa ) \Big(4 \chi _{0,2}^2 \kappa ^2+\chi _{1,1}^2 \kappa
   ^2+4 \chi _{1,3}^2 \kappa ^2+\chi _{2,2}^2 \kappa ^2+4 \chi _{2,3}^2 \kappa ^2-\chi _{1,1} \kappa ^2\\
   &+&2 \chi _{1,1} \chi _{2,2} \kappa ^2-\chi _{2,2} \kappa ^2-24 \chi _{0,2} \chi _{2,3} \kappa +2 \chi _{0,3} \left(-\kappa
   ^2+\left(\kappa ^2-2\right) \chi _{1,1}+\left(\kappa ^2-2\right) \chi _{2,2}+\left(\kappa ^2+4\right) \chi _{3,3}\right) \kappa -7 \chi _{1,1}^2\\
   &-&7 \chi _{2,2}^2-2 \left(\kappa ^2+8\right) \chi _{3,3}^2-2 \chi _{1,1} \chi
   _{2,2}-12 \left(\chi _{1,2}^2-4 \left(\chi _{1,3}^2+\chi _{2,3}^2\right)\right)+\left(2 \kappa ^2-\left(\kappa ^2-16\right) \chi _{1,1}-\left(\kappa ^2-16\right) \chi _{2,2}\right) \chi _{3,3}\Big) \kappa \\
   &+&32 \chi _{1,1}^2+48
   \chi _{1,2}^2-192 \chi _{1,3}^2+32 \chi _{2,2}^2-192 \chi _{2,3}^2+80 \chi _{3,3}^2+16 \chi _{1,1} \chi _{2,2}-80 \chi _{1,1} \chi _{3,3}-80 \chi _{2,2} \chi _{3,3} \Big].
\end{eqnarray*}
\end{widetext}
In both formula $\chi_{i,j}$ for $i\neq j$ correspond to real part of the $\chi$ matrix, the imaginary part that is distinct from the real elements (only present in the first row and column) has no contribution to the variance.

\bibliography{main}

\end{document}